\documentclass[a4paper]{article}
\usepackage{epsfig}
\def\Journal#1#2#3#4{{\em #1} {\bf #2}, #3 (#4) }
\def\NPA{{ Nucl. Phys.} A}
\def\PRL{Phys. Rev. Lett.}

\def\PREP{ Phys. Rep.} 
\def\PRC{{Phys. Rev.} C}
\def\PL {Phys. Lett.}
\def\one{1\!\!1}
\newcommand{\la}{\langle}
\newcommand{\ra}{\rangle}
\newcommand{\tr}{\mbox{Tr}}
\begin{document}
\title{Correlations and the Dirac Structure of the Nucleon Self-Energy}
\author{E. Schiller and  H. M\"{u}ther\\
 Institut f\"{u}r Theoretische Physik\\
Universit\"{a}t T\"{u}bingen, D-72076 T\"{u}bingen, Germany}

\maketitle
\begin{abstract}
The Dirac structure of the nucleon self-energy in symmetric nuclear matter as
well as neutron matter is derived from a realistic meson exchange model for the
nucleon-nucleon (NN) interaction. It is demonstrated that the effects of 
correlations on the effective NN interaction in the nuclear medium can be
parameterized by means of an effective meson exchange. This analysis leads to a
very intuitive interpretation of correlation effects and also provides an
efficient parametrization of an effective interaction to be used in
relativistic structure calculations for finite nuclei.
\end{abstract}

\section{Introduction}
Microscopic studies on the bulk properties of nuclear systems have shown that
two ingredients may be needed to derive nuclear properties from realistic
nucleon-nucleon interactions: the consideration of correlations beyond the
mean-field approach and the relativistic structure of the nucleon self-energy
in the nuclear medium. In fact it is known for many years that attempts to use
realistic NN interactions, i.e.~models for the NN interaction which have been
fitted to the NN scattering phase shifts, in a simple mean field or
Hartree-Fock calculation lead to nuclear systems which are unbound (see
e.g.~the recent review \cite{mupo}). Therefore, various techniques have been
developed to account for correlations beyond the mean field approach including
the Brueckner hole-line expansion\cite{bruek1,bruek2}, the coupled cluster or
``exponential S'' approach\cite{kuem,bish}, the self-consistent evaluation of
Greens functions\cite{wim1}, variational approaches using correlated basis
functions\cite{fhnc1,fhnc2} and recent developments employing quantum
Monte-Carlo techniques\cite{monc1,monc2}. 

In the framework of the Brueckner theory the effects of two-nucleon 
correlations are taken into account by evaluating an effective interaction, 
the so-called $G$-matrix. This $G$-matrix corresponds to the $T$-matrix of NN
scattering, however, in the nuclear medium accounting for Pauli and dispersion
effects. It is obtained by solving the Bethe-Goldstone equation or in the case
of relativistic meson exchange models for the NN interaction by solving an
equation which corresponds to a three-dimensional reduction of the
Bethe-Salpeter equation like the Blankenbecler-Sugar or the Thompson
equation\cite{rupr0}. In the Brueckner-Hartree-Fock (BHF) approximation  the
nucleon self-energy or single-particle potential is then evaluated in terms of
this $G$-matrix. 
 
Accounting for the two-nucleon correlations in this way or by means of other
many-body approaches, one obtains results for the saturation property of
nuclear matter or the binding energy and radius of finite nuclei\cite{mupo},
which are quite reasonable. All such results, however, form the so-called
Coester band\cite{coestba}, i.e. they either predict a binding energy which is to
small or a saturation density which is to large (a to small radius in case of
finite nuclei) as compared to the empirical values. Extensive studies have been
performed to account for three-nucleon correlations\cite{song2}. One finds,
however, that the inclusion of three-nucleon correlations yields a small effect
only, the phenomenon of the Coester band persists for studies within the
framework of Brueckner theory as well as within other approaches to solve the
many-body problem.

It has been suggested to consider three-nucleon forces, which are adjusted to
produce the empirical saturation point of nuclear matter\cite{urba3}.  Another
possibility to shift the calculated saturation point away from the Coester band
towards the empirical value is to account for relativistic effects. These
studies have been motivated by the phenomenology of the Walecka
model\cite{serot}. The NN interaction in this model is described in terms of
the exchange of a scalar meson, $\sigma$, and a vector meson $\omega$.
Calculating the nucleon self-energy, $\Sigma$, from such a meson exchange model
within a Hartree  approximation, one finds that the $\omega$ exchange yields a
component $\Sigma^0$, which transforms under a Lorentz transformation like the
time-like component of a vector, while the scalar meson exchange yields a
contribution $\Sigma^s$, which transforms like a scalar. Inserting this self
energy into the Dirac equation for a nucleon in the medium of nuclear matter
leads to single-particle energies,  which are as small as the empirical value
of -50 MeV. This small binding effect, however, results from a strong
cancellation between the repulsive $\Sigma^0$ and the attractive $\Sigma^s$
component. The attractive scalar component $\Sigma^s$ leads to Dirac spinors
for the nucleons in the nuclear medium, which contain a small component
significantly enhanced as compared to the Dirac spinor of a free nucleon. This
effect is often described in terms of an effective Dirac mass $M^*$ for the
nucleon, which can be of the order of 600 MeV in nuclear matter around
saturation density. This implies that the Dirac spinors for the nucleons in the
nuclear medium are quite different from those in the vacuum. Since these Dirac
spinors are used to evaluate the matrix elements for the meson exchange model of
the NN interaction, this leads to a medium dependence of this interaction, an
effect which influences the calculated saturation property.
 
Also realistic models for the NN interaction contain large contributions from
the exchange of scalar and vector mesons. For such meson exchange potentials,
$V$,   one can determine the Dirac structure of the nucleon self-energy using
the Hartree or Hartree-Fock approximation in a straight forward way. As we have
discussed above, using such realistic NN interactions, one has to account for
correlations beyond the Hartree-Fock approximation and determine the
self-energy in terms of the $G$-matrix rather than the bare interaction $V$,
i.e. perform what is called a Dirac-Brueckner-Hartree-Fock (DBHF) or a
relativistic BHF calculation\cite{anast,brock,malf1,weigel,horst}.
Since, however, this $G$-matrix is obtained as a solution from a
non-relativistic reduction of the scattering equation, it provides matrix
elements only between single-particle states and does not keep track of the
relativistic structure of the effective interaction. 

Various approximation schemes have been developed to determine the Dirac 
structure of the self-energy $\Sigma$ or the structure of the nucleon Dirac
spinors within the context of the DBHF approach. A rather simple scheme has been
suggested by Brockmann and Machleidt\cite{brock}. They determine the momentum 
dependence of the single-particle energy. Identifying this single-particle
spectrum with a corresponding spectrum derived within the Dirac-Hartree
approximation one can extract the effective Dirac mass $M^*$. The underlying
assumption is that the nucleon self-energy is dominated by the scalar,
$\Sigma^s$, and time-like vector component, $\Sigma^0$, which are constants
independent of the momentum of the nucleon. This approximation seems to work
reasonably well for symmetric nuclear matter but it fails in the case of neutron
matter\cite{ulr1}. Therefore calculations of the equation of state for
asymmetric nuclear matter, which are based on this approach\cite{engv1,engv2},
should be considered with some caution.

Another scheme, the so-called projection method, analyses the antisymmetrized
matrix elements of $G$ in terms of sets of operators, which are invariant under
Lorentz transformation. If the relativistic structure of the effective
interaction $G$ is defined in this way, one can derive the Dirac structure of
the self-energy and determine the density dependence of the nucleon
spinors\cite{horo1,boer1}. As will be discussed more in detail below, the
result of this analysis depends on the choice of relativistic invariants. As an
example we mention the well known feature that the one-pion-exchange
contribution yields identical matrix elements for the positive energy Dirac
spinors using pseudo-scalar or pseudo-vector coupling, the Fock contributions to
the scalar part of the self-energy, however, are quite
different\cite{serot,fu1,fu2}. As another example we will discuss the $\rho$
exchange below.

In order to minimise these uncertainties of the projection method, we suggest to
split the $G$ matrix into the Born contribution, the bare interaction $V$, and
the corrections due to the correlations. While the Dirac structure of $V$ is
well defined one may employ the projection method for the correction term
only. It turns out that these corrections can be described rather well in terms
of the exchange of a few effective mesons with high masses, reflecting the short
range of the NN correlation effects, and coupling constants depending on the
nuclear density. The results of this analysis provides some insight into the
effects of NN correlations. This analysis is similar to previous attempts by 
Boersma and Malfliet\cite{boer1} and Elsenhans et al.\cite{elsen}. Because of
its simple structure, however, it might be more appropriate to be used in DBHF
studies of finite nuclei\cite{boer2,fritz} 

After this introduction a brief description of the main ingredients of the
projection technique will be presented in section 2. Results of the analysis for
the effective NN interaction in symmetric nuclear matter and neutron matter
derived from various models of the Bonn potential\cite{rupr0}. After a detailed
discussion of the results in section 3, the main conclusions will be summarised
in section 4.

\section{Relativistic structure of the G-matrix}

The Dirac equation for a nucleon with momentum $k$ in a medium of nuclear matter
can be written 
\begin{equation}
(\not k-M-\Sigma(k))u(\vec{k},s)=0\label{eq:Dirac}
\end{equation}

with the self-energy $\Sigma(k)$ accounting for the mean field generated by the
nuclear medium.
By the requirement of translational and rotational invariance, parity
conservation and time reversal invariance, the general form of the Dirac
structure of the self-energy is given in the nuclear matter rest frame 
as $4\times 4$-Matrix by
\begin{equation}
\Sigma(k)=\Sigma^s(k)-\gamma^0\Sigma^0(k)+\vec{\gamma}\cdot\vec{k}\Sigma^v(k)
\label{eq:Sigma}
\end{equation}
where $\Sigma^s, \Sigma^0, \Sigma^v$ are functions, depending for on-shell
nucleons ($k^0=E(\vec{k})$) only on the absolute value of the three-momentum
$k\equiv|\vec{k}|$ and the Fermi-momentum $k_f$, which is related to the
density via $\rho=\delta/(3\pi^2)k_f^3$ where the isospin degeneracy yields
$\delta=2$ for  nuclear matter and $\delta=1$ for neutron matter. The density
dependence will be suppressed throughout this section.

The components of the self-energy are easily determined by taking traces of the
form
\begin{eqnarray}
\Sigma ^{s}(k)=\frac{1}{4}\tr[\Sigma(k) ]\,, \qquad &
\Sigma ^{0}(k)=-\frac{1}{4}\tr[\gamma ^{0}\Sigma(k)]\,, &\qquad
\Sigma ^{v}(k)=-\frac{1}{4}\tr[\vec{\gamma}\cdot \hat{k}\Sigma(k)]
\label{eq:DecomSelf}
\end{eqnarray}
Introducing the effective quantities
\begin{eqnarray}
M^*(k)=\frac{M+\Sigma^s(k)}{1+\Sigma^v(k)} \qquad &
\qquad E^*(k)=\frac{E+\Sigma^0(k)}{1+\Sigma^v(k)}\label{eq:effm}
\end{eqnarray}
the Dirac equation can be rewritten in compact form 
setting
$\vec{k^*}=\vec{k}$ 
\begin{equation}
(\not k^* - M^*)u(\vec{k},s)=0\label{eq:DiracEff}
\end{equation}
which is formal identical to the Dirac equation in the vacuum case. 
Therefore the positive energy solution to 
eq.~(\ref{eq:DiracEff}) is given as
\begin{equation}
u(\vec{k},s)=\sqrt{\frac{E^*(k)+M^*(k)}{2M^*(k)}}
\left(\begin{array}{c}
1 \\ 
\frac{\vec{\sigma}\cdot\vec{k}}{(E^*(k)+M^*(k))}
\end{array}
\right)
\chi_s
\label{eq:Spinor}
\end{equation}
with the covariant normalisation
\begin{eqnarray}
\overline{u}u=1\,,\qquad& \qquad u^+u=\frac{E^*(k)}{M^*(k)}
\end{eqnarray}
and the in medium on-shell relation
\begin{equation}
E^{*2}(k)=M^{*2}(k)+\vec{k}^2
\end{equation}

The self-energy itself is connected to the two-particle effective interaction in
the medium, the G-matrix, through the Hartree-Fock relation
\begin{eqnarray}
\Sigma_{\alpha\beta}(k)&=&-i\int_{\vec{q}\le k_f}\frac{d^4q}{(2\pi)^4}
\tilde{g}_{\tau\sigma}(q)  
\left\{G(k,q;k,q)_{\alpha\sigma;\beta\tau}-G(k,q;q,k)_{\alpha\sigma;\tau\beta}
\right\}\nonumber \\
&=&-i \int_{\vec{q}\le k_f}\frac{d^4q}{(2\pi)^4} \tilde{g}_{\tau\sigma}(q) 
G^A(k,q;k,q)_{\alpha\sigma;\beta\tau}\nonumber\\
&=&-i \int_{\vec{q}\le k_f}\frac{d^4q}{(2\pi)^4} 
\tr[\tilde{g}(q)G^A(k,q;k,q)]_{\alpha\beta}
\label{eq:SelfOp}
\end{eqnarray}
with the propagator for real nucleons propagating on-shell inside the
Fermi-sea in the nuclear matter rest frame
\begin{equation}
\tilde{g}_{\tau\sigma}(q)=\frac{i\pi}{E^*(q)}(\not q^*+M^*)_{\tau\sigma}
\delta(q^0-E(q))\Theta(k_f-|\vec{q}|)
\end{equation}

The equation above (\ref{eq:SelfOp}) must be understood as an operator equation.
Therefore, also the $G$-matrix
is needed in terms of Dirac operators as will be discussed now.

One of the main ingredients of the DBHF approach is the derivation of the
effective interaction $G$ in the medium from a realistic nucleon-nucleon
interaction $V$. This is achieved by solving the relativistic Bethe-Goldstone
equation, a kind of Thomson equation for the scattering of two nucleons in
nuclear matter 
\begin{equation}
G(\vec{q'},\vec{q}|\vec{P},\omega)=
V(\vec{q'},\vec{q}) +
 \int\frac{d^3k}{(2\pi)^3}
 V(\vec{q'},\vec{k})
\frac{M^{*2}}{E^{*2}_{\vec{P}+\vec{k}}} 
\frac{Q(\vec{k},\vec{P})}{\omega-2E^{*}_{\vec{P}+\vec{k}}}
G(\vec{k},\vec{q}|\vec{P},\omega)
\label{eq:BGeq}
\end{equation}
with the starting energy
\begin{equation}
\omega =2 E^*_{\vec{P}+\vec{q}}
\end{equation}
where $\vec{P}=\frac{1}{2}(\vec{p}_1+\vec{p}_2)$ denotes the center of mass
momentum and $\vec{q}=\frac{1}{2}(\vec{p}_1-\vec{p}_2)$ is the relative
momentum of the  initial state, $\vec{k}$ and $\vec{q'}$ are the relative
momenta of the intermediate and final states of two interacting nucleons. The
accessible intermediate states are restricted by the Pauli operator $Q$ to
momenta above the Fermi momentum $k_f$.  To shorten the notation isospin indices
will be suppressed throughout this section. 

To account for the spin, the $G$-matrix
$\la\lambda_1'\lambda_2'|G(\vec{q'},\vec{q})|\lambda_1\lambda_2\ra$ is
conveniently obtained in a basis of helicity states, where $\lambda_1', \lambda_2'$
and $\lambda_1, \lambda_2$ are the helicities of the final and initial state of
particle $1$ and $2$.
Demanding parity conservation, time
reversal invariance and conservation of total spin reduces the $16$
possible helicity amplitudes to six independent amplitudes, which are
normally chosen to be
\begin{eqnarray}
G_1 = \la ++|G(\vec{q'},\vec{q})|++ \ra\, , \qquad & \qquad
G_2 = \la ++|G(\vec{q'},\vec{q})|-- \ra\, , \nonumber\\
G_3 = \la +-|G(\vec{q'},\vec{q})|+- \ra \, , \qquad &  \qquad
G_4 = \la +-|G(\vec{q'},\vec{q})|-+ \ra\, ,  \nonumber\\
G_5 = \la ++|G(\vec{q'},\vec{q})|+- \ra\, , \qquad  & \qquad
G_6 = \la +-|G(\vec{q'},\vec{q})|++ \ra\, , 
\end{eqnarray}
where $+$, $-$ denotes the helicity $\frac{1}{2}$ and $-\frac{1}{2}$,
respectively. The
explicit dependence on $\vec{P}$ and $\omega$ has been omitted.
The number of independent helicity matrix elements is further reduced to five 
in the case of
on-shell scattering, $|\vec{q'}|=|\vec{q}|$, which implies
\begin{equation}
G_5 = - G_6
\end{equation}

In order to evaluate the nucleon self-energy eq.(\ref{eq:SelfOp}), the
operator structure of the $G$-Matrix is needed for the on-shell case in the
nuclear matter rest frame; on the other hand the determination of the operator
structure 
is most easily done in the center of mass frame where $\vec{P}=0$ and the scattering
angle $\vartheta$ between the relative momenta $\vec{q'}$, $\vec{q}$ is
fixed at $\vartheta=0$. Therefore, only $q\equiv|\vec{q}|$ have to be
considered in the following as argument.
The operator structure is then given by expanding the
$G$-matrix in terms of five independent Fermi covariants according to
the five helicity matrix elements of the $G$-matrix:
\newcommand{\argo}{q}
\begin{eqnarray}
{
\la\lambda_{1}',\lambda_{2}'|
G(\argo)|
\lambda_{1},\lambda_{2}\ra_{A}} & = &
\sum_{i=S,V,T,A,PS} \Gamma^{i}_{D}(\argo)\;
\langle\ |
\hat{T}^{(1)}_{i}\hat{T}^{(2)}_{i}|\:
\ra_{D}  - \Gamma^{i}_{X}(\argo)\;
\langle\ |
\hat{T}^{(1)}_{i}\hat{T}^{(2)}_{i}|\:
\ra_{X} 
\nonumber\\[2ex]
& = &\sum_{i=S,V,T,A,PS}\Gamma^{i}_{A}(\argo)\;
\langle\lambda_{1}',\lambda_{2}',\vec{q}|
\hat{T}^{(1)}_{i}\hat{T}^{(2)}_{i}|\:
\vec{q},\lambda_{1},\lambda_{2}\ra_{D}
\label{eq:DecomG}
\end{eqnarray}
with the Fermi covariants
\begin{equation}
\hat{T}_{i}\in\{1,\gamma^{\mu},\sigma^{\mu\nu},\gamma^{5}\gamma^{\mu},
\gamma^{5}\}
\label{eq:FermiCo}
\end{equation}
and the Lorentz invariant amplitudes
\begin{equation}
\Gamma^{i}_{A}(\argo)=\Gamma^{i}_{D}(\argo) -
\sum_{k=1}^{5} F_{ki}\Gamma^{k}_{X}(\argo)
\end{equation}
where the $F_{ki}$ is the well known Fierz-Transformation
\begin{equation}
(F)_{ki}=\frac{1}{4}\left(\begin{array}{ccccc}
  1 & -1 & -\frac{1}{2} &  1 & -1 \\
 -4 & -2 &            0 & -2 & -4 \\
-12 &  0 &           -2 &  0 & 12 \\
  4 & -2 &            0 & -2 &  4 \\
 -1 & -1 &  \frac{1}{2} &  1 &  1 
\end{array}\right)\label{eq:Fierz}
\end{equation}

The subscript $A$ in eq.(\ref{eq:DecomG}) indicates, that only
antisymmetrized matrix elements are 
obtained as solution of the Bethe-Goldstone equation and therefore only those can
be analysed. Here, the explicit splitting of the antisymmetrized matrix element in its
direct and exchange part (labelled with the subscripts $D$ and $X$) only
illustrates, that, using the local Fermi 
covariants eq.(\ref{eq:FermiCo}), both parts can be rewritten with the help of 
the Fierz-Transformation in terms of the antisymmetrized amplitudes $\Gamma_A^i$
and the direct matrix element of the Fermi-covariants. Nevertheless, the
direct and exchange part cannot be determined from the antisymmetrized
matrix elements and it is not required to evaluate the self-energy.

Inverting eq.(\ref{eq:DecomG}) yields the antisymmetrized amplitudes
$\Gamma^i_{A}$, 
where only the scalar $\Gamma^{S}_{A}$ and vector $\Gamma^{V}_{A}$
amplitudes are needed to evaluate the self-energy using eq(\ref{eq:SelfOp}).

\begin{eqnarray}
\Sigma^{s}(k_1)&=&-\frac{1}{2\pi^{2}}\int_{0}^{k_f}d^{3}\vec{k_2}\
\frac{M^{*}}{E^{*}} \;\Gamma^{S}_{A}(q)\nonumber\\[1.5ex]
\Sigma^{0}(k_1)&=&-\frac{1}{2\pi^{2}}\int_{0}^{k_f}d^{3}\vec{k_2}
\ \Gamma^{V}_{A}(q)\nonumber\\[1.5ex]
\Sigma^{v}(k_1)&=&-\frac{1}{2\pi^{2}}\int_{0}^{k_f}d^{3}\vec{k_2}\ 
\frac{{k_2}^{\:*}}{E^{*}}\;\Gamma^{V}_{A}(q)\label{eq:selfgam}
\end{eqnarray}
Here, $k_1$ and $k_2$ are the single particle momenta in the nuclear matter
rest frame. The relative momentum $q$, defined in the c.m. frame, is related
with $k_1$ and $k_2$ via $q=\sqrt{s/4-M^{*2}}$ where
$s=(E^*(k_1)+E^*(k_2))^2-(\vec{k_1}+\vec{k_2})^2$ is the invariant mass. This
transformation between the nuclear matter rest frame and the center of mass
frame is described in detail in\cite{horo1}.

The used set of Fermi-covariants is sufficient to reproduce the matrix elements
of the effective interaction $G$ for the positive energy solutions of the Dirac
equations. From the resulting amplitudes $\Gamma^{S}_{A}(q)$ and
$\Gamma^{V}_{A}(q)$ one can immediately determine the components of the
self-energy according to (\ref{eq:selfgam}).  Nevertheless this procedure is
not unique and depends on the chosen set of operators $\hat T$.  A well known
example which illustrates this dependence  is the pion-exchange part of the NN
interaction, which either is described by a  pseudo-scalar or a pseudo-vector
coupling. Both couplings yields for on-shell nucleons the same  matrix
elements, if the coupling constants $f_{pv}$ and $g_{ps}$ obey the relation
$f_{pv}/m_{\pi}=g_{ps}/(2M)$, where $m_{\pi}$ is the pion-mass and $M$ the mass
of the nucleon. Nevertheless, the components of the self-energy are complete
different for both couplings. A more detailed discussion of this and other 
examples will be given in the next section. 

To circumvent aforementioned problem, which already occurs in the case if
the effective interaction $G$ is replaced with the bare nucleon-nucleon
interaction $V$, the $G$-matrix is split into two parts
\begin{equation}
G=V+\triangle G \label{eq:v+delG}
\end{equation}
The decomposition eq.(\ref{eq:DecomG}) is applied only to the residual
part $\triangle G$ since the explicit Dirac structure of $V$ and the way to
evaluate the components of the self-energy from $V$ is known explicitly.

\begin{figure}[b]
\begin{center}
\epsfig{file=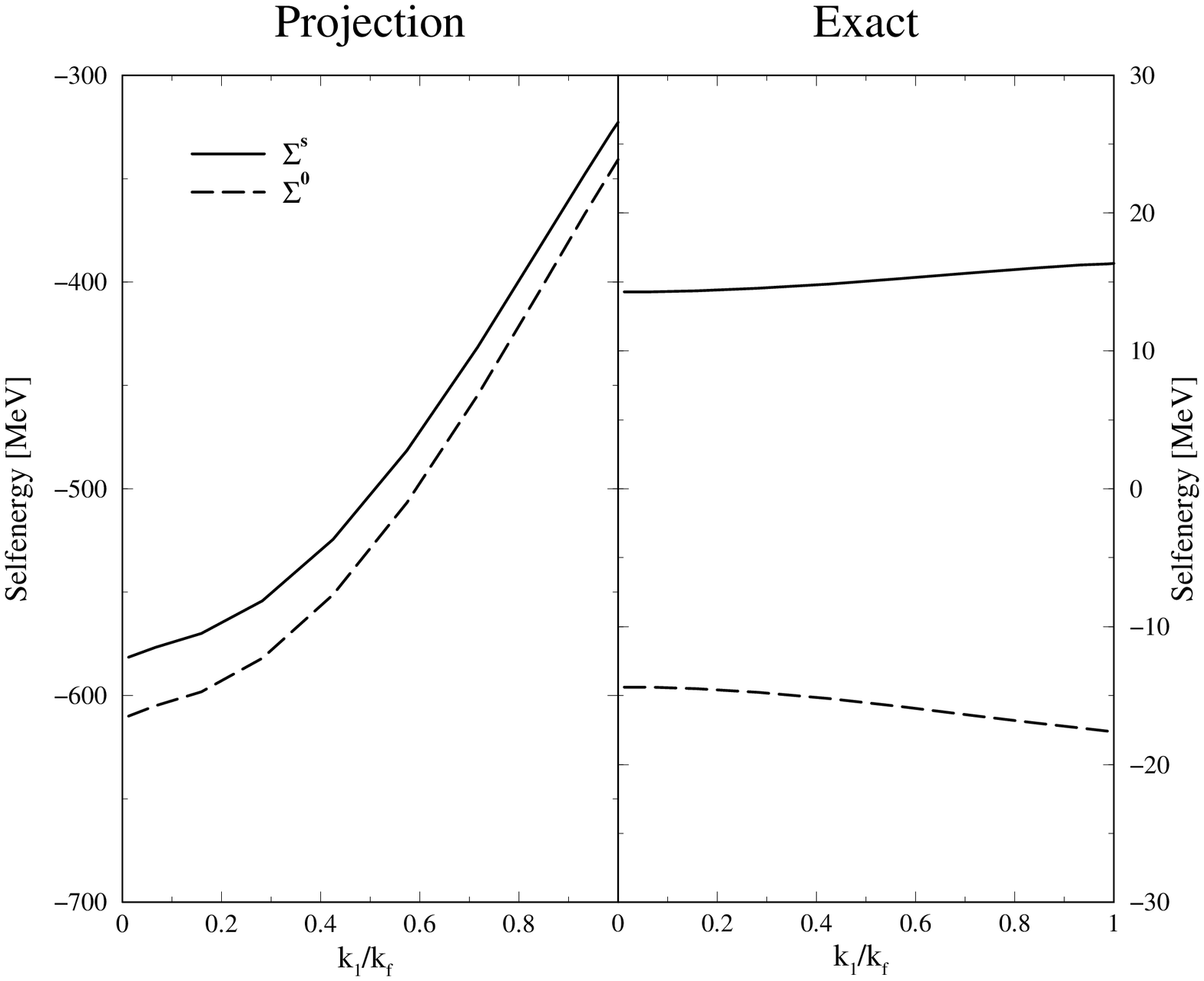,width=9cm}
\end{center}
\caption{\label{fig:pion} Contribution of the $\pi$ exchange to the scalar 
($\Sigma^s$) and vector component ($\Sigma^0$) of the nucleon self-energy in
nuclear matter with $k_F$ = 1.36 fm$^{-1}$. Results derived from the projection
method (pseudo-scalar coupling) are displayed in the left part of the figure,
while those derived from the original Bonn interaction (pseudo-vector coupling)
are given in the right part. Note the different scales in the two parts of the
figure.}
\end{figure}

\section{Results and discussion}
In the first part of this section we would like to show some ambiguities of the
projection method which is related to the choice of the covariant operators in
(\ref{eq:FermiCo}). For that purpose we consider various components of the 
bare NN interaction $V$ using as an example the OBE potential Bonn B as defined 
in table A.2 of \cite{rupr0}. If one considers only the $\pi$-exchange 
contribution and analyses the matrix elements of $V_{\pi}$ by means of the
projection formalism discussed in section 2, one obtains large components for
Lorentz invariant scalar and vector amplitudes $\Gamma^{s}_{A}$ and
$\Gamma^{v}_{A}$ which are due to the Fock-exchange terms of the pseudo-scalar
operator representing the $\pi$ exchange in the projection formalism.
Integrating these amplitudes as indicated in (\ref{eq:selfgam}) leads to
large scalar ($\Sigma^s$) and time-like vector components ($\Sigma^0$).

This is visualised in the left part of Fig.~\ref{fig:pion}, where these
components calculated for symmetric nuclear matter at the empirical saturation
density (Fermi momentum $k_F$ = 1.36 fm$^{-1}$)  are presented as a function of 
the nucleon momentum $k_1$. Note that this analysis yields an absolute value for
the scalar component which gets as large as 600 MeV, which implies that the
effective Dirac mass, $M^*$, of the nucleon in the nuclear medium would become 
as small as 340 MeV. Furthermore this analysis yields quite a strong momentum
dependence of the scalar and vector components of the self-energy $\Sigma$.
Because of the strong cancellation between the scalar and vector components, the
total effect of the $\pi$-exchange on the single-particle energy  
\begin{equation}
\epsilon (k) = \sqrt{M^{*2}(k)+\vec{k}^2} - \Sigma^0 (k) - M \label{eq:epsik}
\end{equation}
is rather small. 

Calculating the self-energy directly from the $\pi$-exchange part in the Bonn B
potential leads to quite different results as one can see from the right-hand
side of Fig.~\ref{fig:pion}. In this case the scalar part of the self-energy
has the opposite sign, it is small (of the order of 15 MeV) and exhibits a very
weak momentum dependence. The single-particle energy for the nucleons in
nuclear matter, calculated according to (\ref{eq:epsik}), is identical to the
one derived from the projection method, the Dirac structure of the self-energy,
however, is completely different. The reason for this difference is well known:
The matrix elements of the $\pi$-exchange potential, calculated for the 
positive energy solution of the Dirac equation for the nucleon, are independent
of assuming either pseudo-scalar or pseudo-vector coupling for the $\pi$N vertex.
The coupling between the positive and negative energy spinors, however, is quite
different. The projection method analyses the matrix elements of the positive
energy solutions in term of a pseudo-scalar operator, leading to large components
of the self-energy. Assuming a pseudo-vector coupling for the pion, as it is 
done in the original NN interaction $V$ yields much smaller components in the
self-energy. This feature is known for a long time (see e.g.~\cite{serot}). It
is the origin of the strong momentum dependence of the self-energy observed in
\cite{fu1}, where the projection method has been applied. The problem might be 
cured by analysing the interaction in terms of the pseudo-scalar operator,
transforming the pseudo-scalar component into a corresponding pseudo-vector term
and evaluate the self-energy with this pseudo-vector part (see
e.g.\cite{boer1,fu2}). 

\begin{figure}[b]
\begin{center}
\epsfig{file=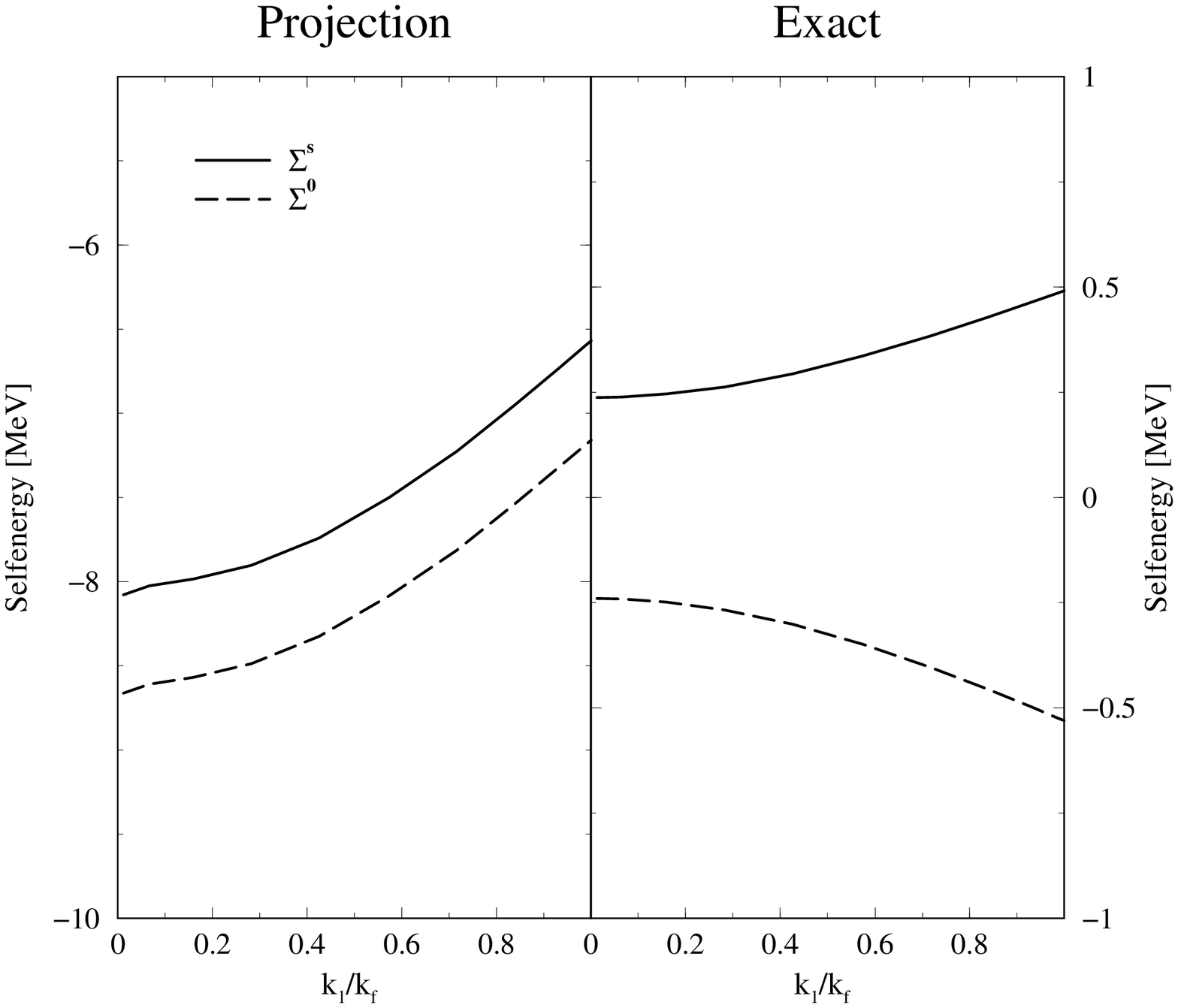,width=9cm}
\end{center}
\caption{\label{fig:eta} Contribution of the $\eta$ exchange to the scalar 
and vector component of the nucleon self-energy in
nuclear matter. Further details see figure \protect{\ref{fig:pion}}.} 
\end{figure}

Such ambiguities of the projection method, however, not only arise for the
$\pi$-exchange part. Another example is of course the exchange of the
pseudo-scalar - isoscalar meson, the $\eta$ meson. Because of its weaker coupling
constant and its higher mass, it does not play such a significant role as the
$\pi$-exchange. Nevertheless, also in this case the differences between the
self-energy components derived via the projection method and from the direct
evaluation are non-negligible as one can see from Fig.~\ref{fig:eta}. 

\begin{figure}[tb]
\begin{center}
\epsfig{file=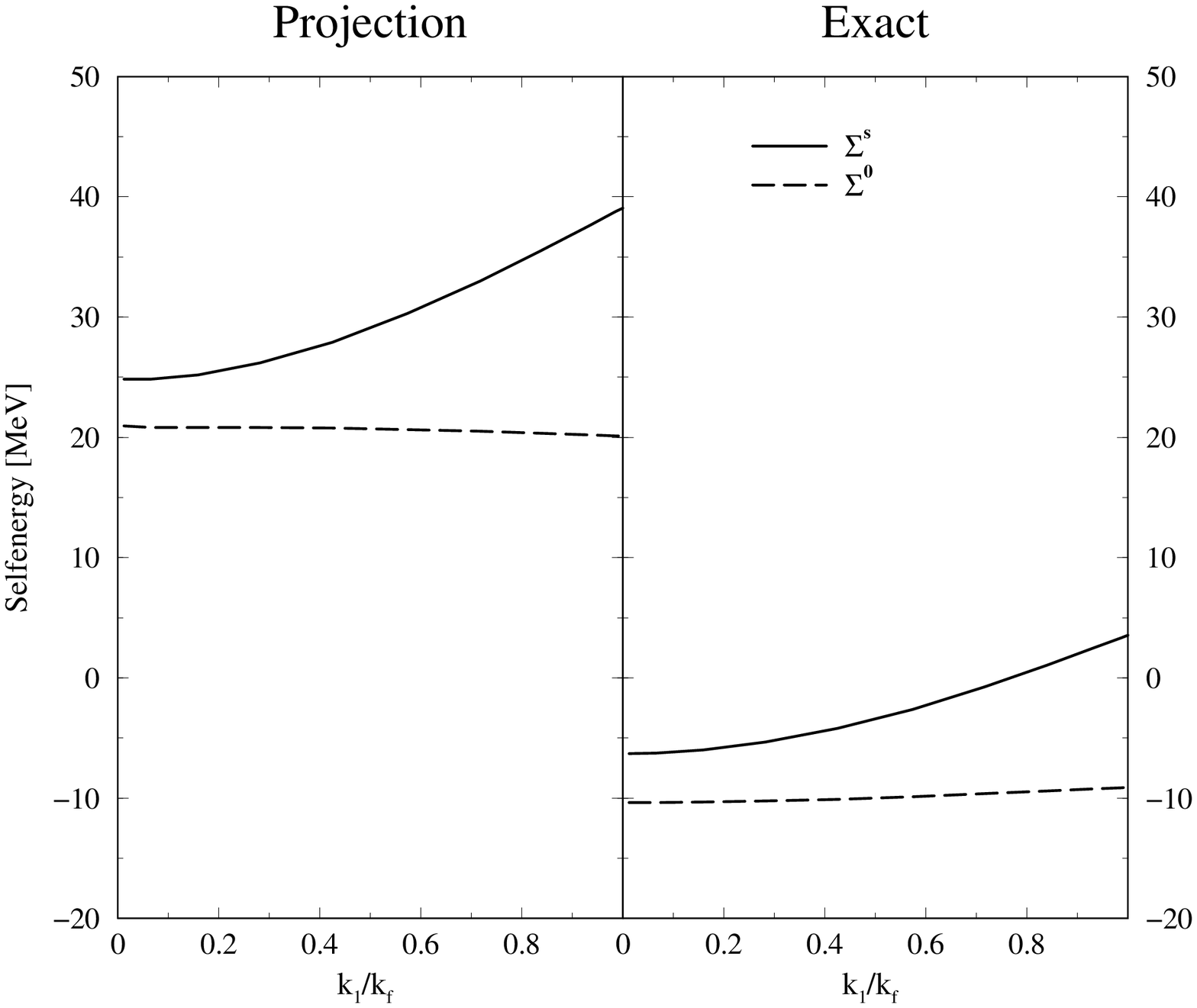,width=9cm}
\end{center}
\caption{\label{fig:rho} Contribution of the $\rho$ exchange to the scalar 
and vector component of the nucleon self-energy in
nuclear matter. Further details see figure \protect{\ref{fig:pion}}.} 
\end{figure}

The $\rho$-exchange contribution to $V$ is considered as a last example for
such a comparison between the self-energies calculated via the projection
method and the direct evaluation of the meson exchange term. Results are
displayed in Fig.~\ref{fig:rho}. The original $\rho$-exchange term of the Bonn
B potential contains a vector coupling but also a strong tensor coupling
(Pauli-coupling) term in the Lagrangian for the $\rho$N interaction. Also these
two different coupling modes cannot be resolved by means of the projection
method. The difference between the direct evaluation of the self-energy and
using the projection formalism are remarkable also in this case. 

Therefore, in  order to minimise the ambiguities in the projection method, we
suggest to split the effective interaction $G$ according to eq.(\ref{eq:v+delG})
into the bare interaction $V$ and the correction term $\Delta G$, representing
the corrections which are due to the correlations. The Dirac structure of the
bare interaction is directly known, and the projection method has to be applied
to the analysis of $\Delta G$ only. This scheme is advantageous also from
another point of view: One must keep in mind that the matrix elements of $V$ and
$G$ in the helicity basis are dominated by the one-pion-exchange contribution.
Note that the ratio of coupling constant and mass of the meson,
$g_\alpha^2/m_{\alpha}^2$, which is a measure for the importance of the various
meson exchange contribution is about a factor 25 larger for $\alpha = \pi$ than
for the $\sigma$ or $\omega$ meson. Analysing $\Delta G$ this dominating
$\pi$-exchange contribution, including the momentum dependencies, which are
related to the form-factors for the $\pi$N vertex, are removed, which stabilises
the numerical analysis significantly.

\begin{figure}[tb]
\begin{center}
\epsfig{file=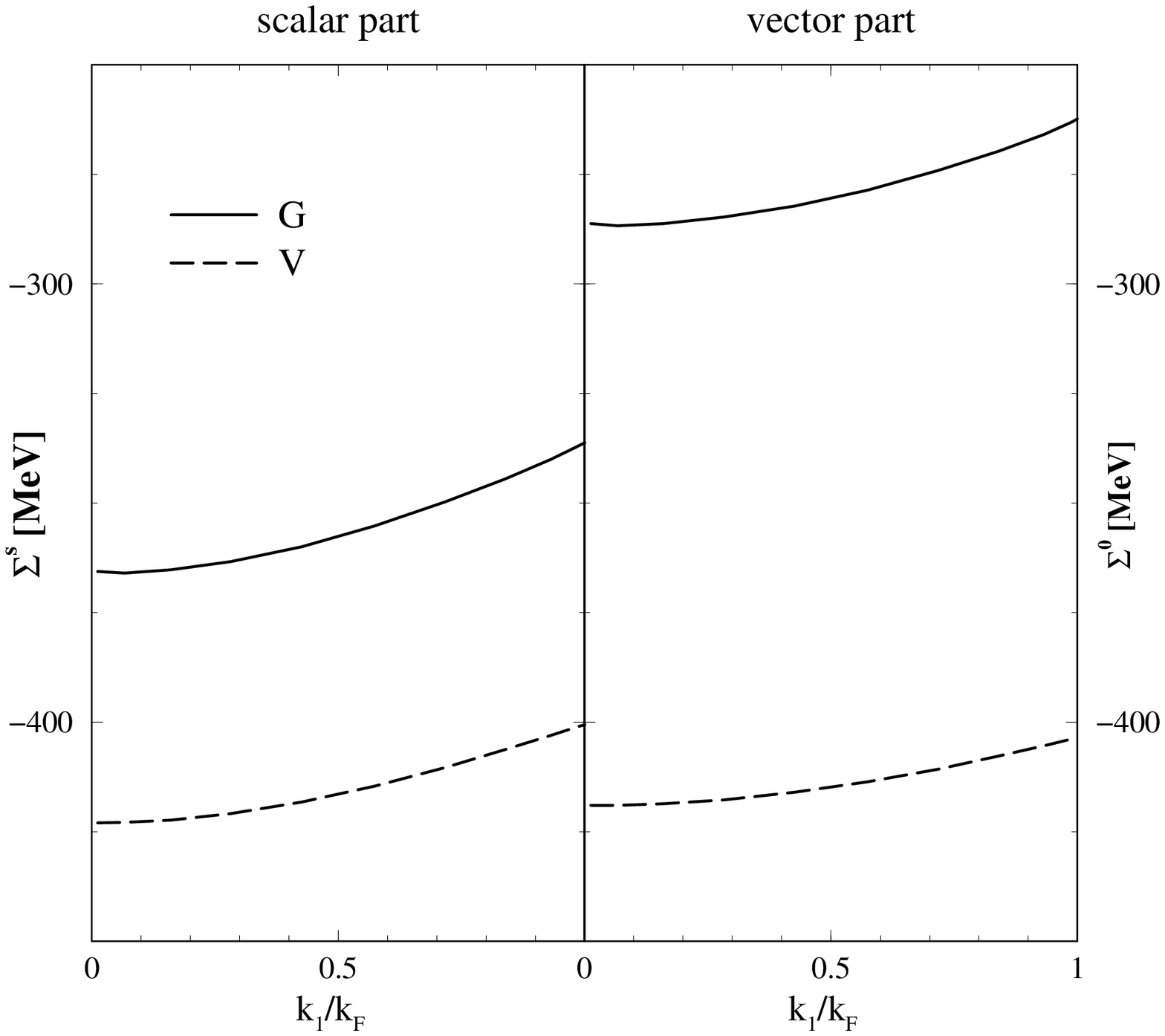,width=9cm}
\end{center}
\caption{\label{fig:gv136} The scalar ($\Sigma^s$, left part) and time-like 
vector component ($\Sigma^0$, right part) of the nucleon self-energy in
nuclear matter at $k_F$ = 1.36 fm$^{-1}$. Results obtained for the bare NN
interaction $V$ (Bonn B potential of \protect\cite{rupr0}, dashed lines) are
compared to those derived from the $G$-matrix (solid lines).} 
\end{figure}

Results for the nucleon self-energy calculated again for symmetric nuclear
matter at a Fermi momentum $k_F$ = 1.36 fm$^{-1}$ are displayed in
Fig.~\ref{fig:gv136}. The contribution of the bare NN interaction $V$ to the 
scalar and vector part of the self-energy (see dashed lines in both parts of the
figure) yields rather similar values ranging from -420 MeV to -400 MeV for
$k_1/k_F$ between 0 and 1. This implies that the single-particle energies tend
to be positive and one obtains no binding energy for nuclear matter. This is in
line with the non-relativistic studies mentioned in the introduction: Using
realistic NN interactions one does not obtain any binding energy if correlations
beyond the Hartree-Fock approach are ignored. 

Adding the contributions of $\Delta G$ to the various components of
the self-energy, one obtains the results displayed by the solid lines in
Fig.~\ref{fig:gv136}. The correlation effects contained in $\Delta G$ reduce the
absolute value for the scalar as well as the vector component of the
self-energy. One could argue that the short-range correlations lead to a
reduction of the wave function at small relative distances. This quenching of
the relative wave function reduces the effects of the $\sigma$ exchange, which
is the dominating contribution to $\Sigma^s$, as well as the $\omega$ exchange,
which is the driving term in $\Sigma^0$. Since the $\omega$ exchange is of
shorter range than the $\sigma$ exchange, the mass of the $\omega$ meson
($m_\omega$ = 783 MeV) is larger than the one of the $\sigma$ ($m_\sigma$ = 550
MeV), the quenching of the short-range components in the relative wave function
is more important for the $\omega$ than for the $\sigma$ exchange. This explains
that the absolute value of $\Sigma^0$ is reduced stronger by $\Delta G$ than the
absolute value of $\Sigma^s$. This difference on the effect of correlations on
$\Sigma^s$ and $\Sigma^0$ leads to attractive single-particle energies and
binding energy of nuclear matter.

\begin{figure}[tb]
\begin{center}
\epsfig{file=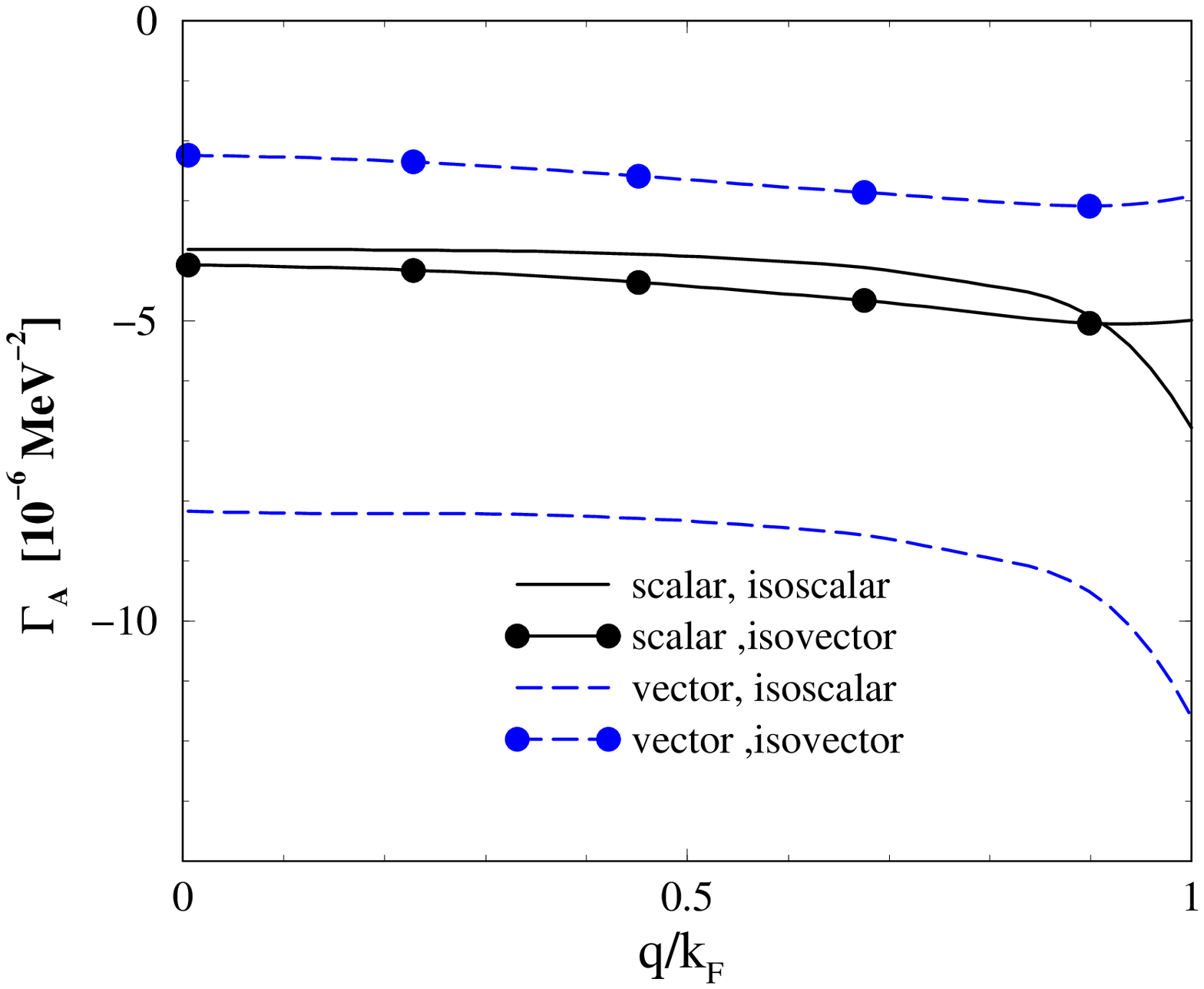,width=9cm}
\end{center}
\caption{\label{fig:amp2} Antisymmetrized interaction amplitudes $\Gamma_A^i$,
derived from the analysis of $\Delta G$ presented as a function of the relative
momentum of the interacting nucleons in their center of mass frame} 
\end{figure}

Its worth noting that the inclusion of $\Delta G$ leads to a constant shift in
the components of the self-energy, which is almost independent of the momentum
of the nucleon. This very weak momentum dependence of the $\Delta G$ effects
can also bee seen from Fig.~\ref{fig:amp2}, where the antisymmetrized
amplitudes $\Gamma_A^i$ (see eq.(\ref{eq:DecomG})) are displayed. These
amplitudes are derived from the analysis of $\Delta G$. Also these
interaction amplitudes are almost independent of the relative momentum $q$.
Deviations from the constant value are observed only for $q \to k_F$, i.e.~for
momenta for which the  energy denominator in the Bethe-Goldstone equation
approaches the pole.

\begin{figure}[tb]
\begin{center}
\epsfig{file=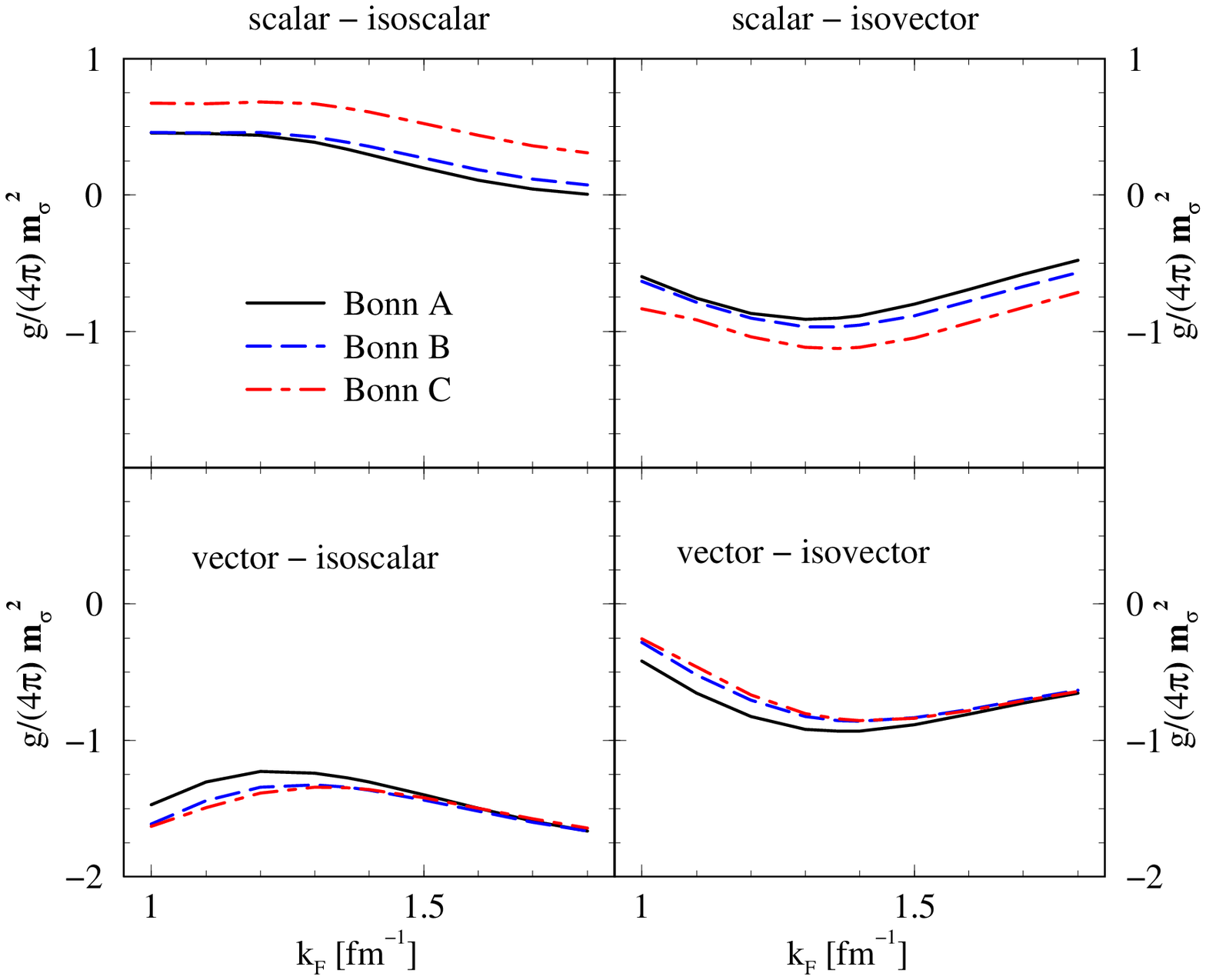,width=12cm}
\end{center}
\caption{\label{fig:coudG} Coupling constants for the parametrization of
$\Delta G$ according to eq.(\protect\ref{eq:pardelg}) for the three versions of
the Bonn potentials. These coupling constants
are presented as a function of the Fermi momentum $k_F$ of symmetric nuclear
matter. The values of the constants are multiplied with the square of the mass 
of the $\sigma$ meson ($m_\sigma = 550$ MeV) and divided by $4\pi$
to obtain dimensionless quantities of the order 1.} 
\end{figure}

This observation suggests to parameterize the component $\Delta G$ in terms of
the exchange of effective mesons with infinite mass. To put it in different
words: The correlation effects contained in $\Delta G$ are described in terms of
an effective interaction with zero range. If we focus the attention on the
scalar and vector interaction amplitudes only, at each density 4 coupling
constants are required to parameterize $\Delta G$ in the form
\begin{equation}
\Delta G = \left[g_{s,s} \one \one  + g_{v,s} \gamma^\mu \gamma_\mu \right]  +
\left[g_{s,v} \one \one  + g_{v,v} \gamma^\mu \gamma_\mu \right] \vec\tau \cdot
\vec \tau \label{eq:pardelg}\, .
\end{equation}
The results for these effective coupling constants are displayed in
Fig.~\ref{fig:coudG}.

The value of these coupling constants indicate that $\Delta G$ is in general
weaker than $V$ but yields a non-negligible correction to the bare interaction
$V$. The parameters are rather similar for the three versions of the Bonn
potential (Bonn A, B and C as defined in table A.2 of \cite{rupr0}). The
absolute values are typically larger for Bonn C than for the other two
reflecting the fact that this potential yields slightly stronger correlation
effects. The density dependence of these parameters is weak but non-negligible.
The largest absolute values are observed for vector - isoscalar pseudo meson
($g_{v,s}$). This reflects the fact, which we mentioned already above, that
correlations yield a suppression of the $\omega$ meson exchange in particular.

\begin{figure}[tb]
\begin{center}
\epsfig{file=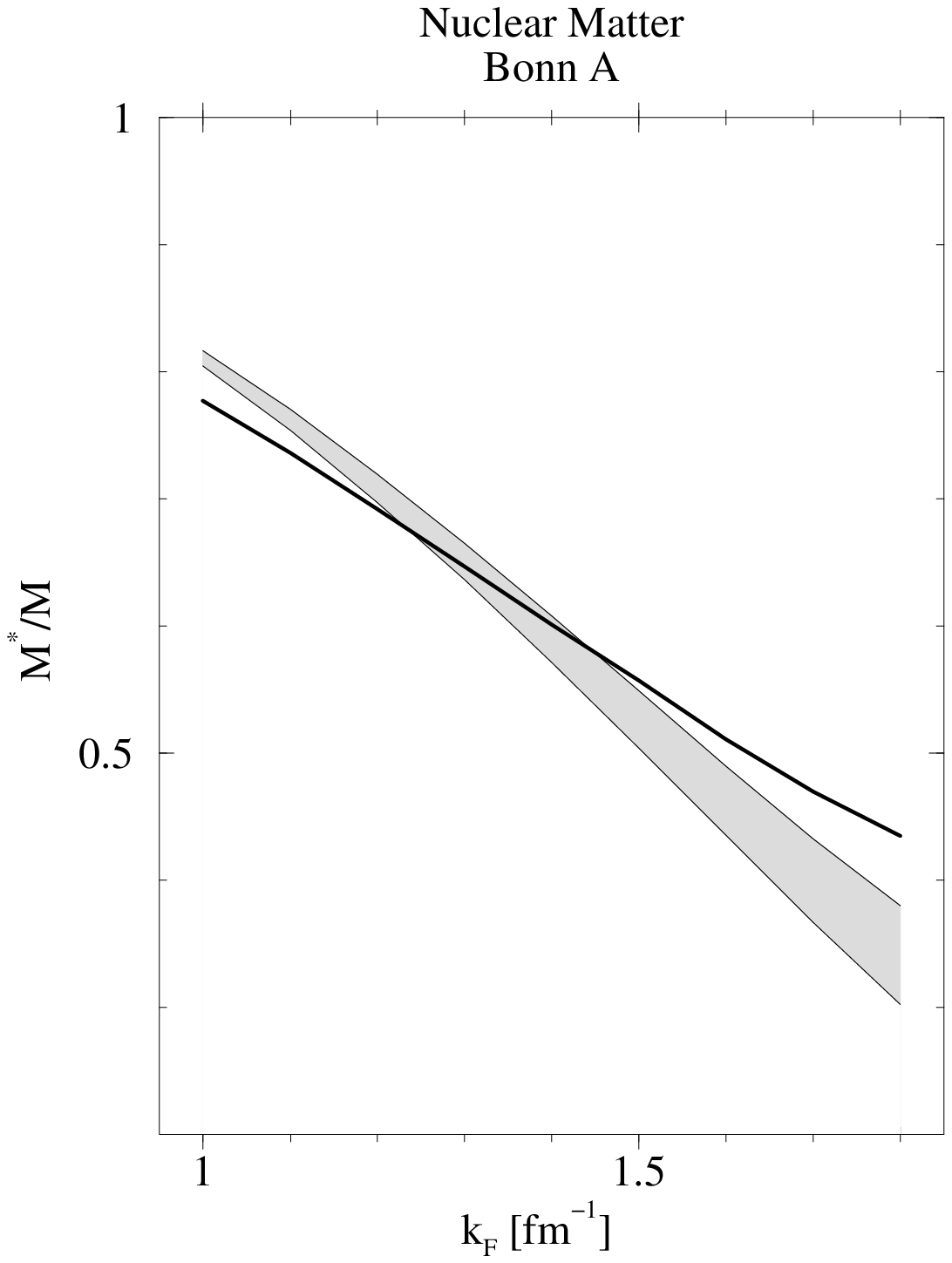,width=6cm}
\end{center}
\caption{\label{fig:mskma} Effective mass $M^*$ characterising the Dirac spinor
of the nucleon in nuclear matter as a function of the Fermi momentum $k_F$. The
range of the momentum-dependent masses derived from the projection method are
indicated by the shaded area, the result obtained from an analysis of the
single-particle spectrum is displayed by the solid line. The data have been
derived from the Bonn A potential.}
\end{figure}

\begin{figure}[tb]
\begin{center}
\epsfig{file=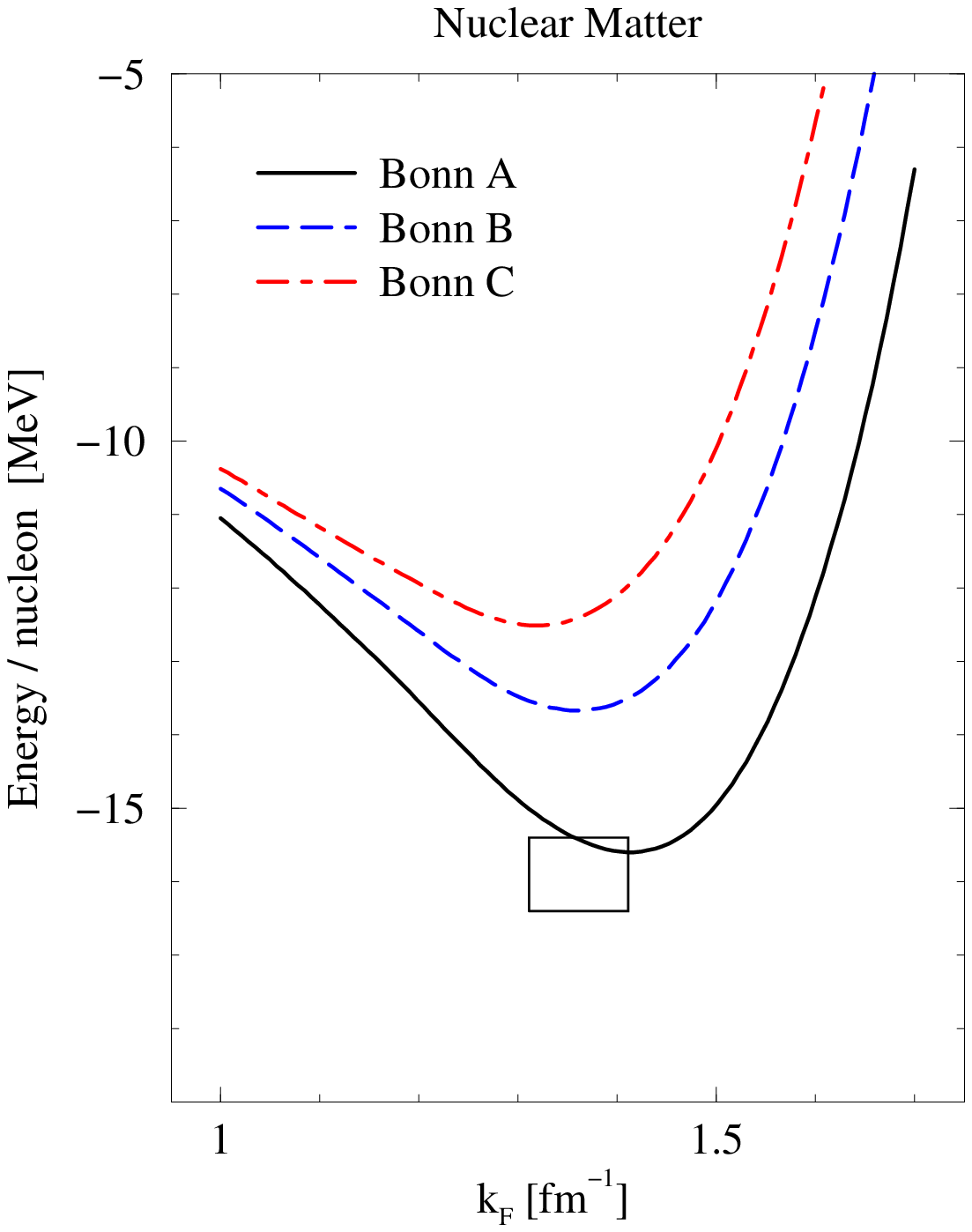,width=6cm}
\end{center}
\caption{\label{fig:ebkm} Calculated binding energy of nuclear matter as a
function of the Fermi momentum $k_F$ using the three different versions of the
Bonn potential}
\end{figure}

Results for the properties of symmetric nuclear matter are displayed in
Figs.~\ref{fig:mskma} and \ref{fig:ebkm}. Fig.~\ref{fig:mskma} displays the
density dependence of the effective Dirac mass $M^*(k)$ defined in 
eq.~(\ref{eq:effm}). The
shaded area at each Fermi momentum $k_F$ indicates the range of values for
$M^*(k)$, which is obtained at the corresponding density using the projection
method. For a comparison we also present the effective mass derived from the
momentum dependence of the single-particle energy as proposed by  Brockmann 
and Machleidt\cite{brock}. The analysis of the single-particle energy, which is
much simpler than the projection scheme, yields rather similar results for 
symmetric nuclear matter. Therefore also the calculated binding energies, 
derived from the projection scheme (see Fig.~\ref{fig:ebkm}) are very close to
those obtained in \cite{brock}.

\begin{figure}[tb]
\begin{center}
\epsfig{file=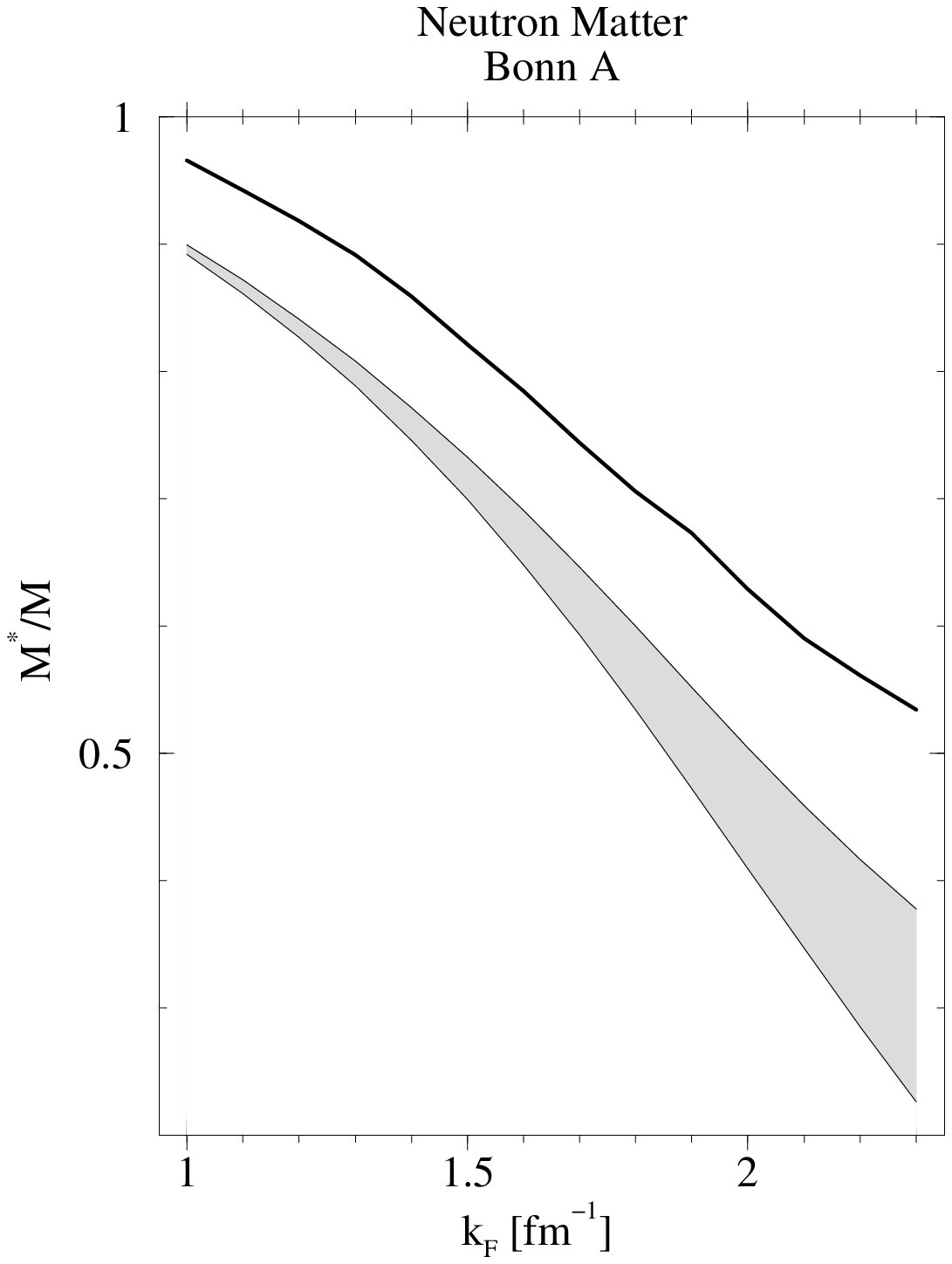,width=6cm}
\end{center}
\caption{\label{fig:msnma} Effective mass $M^*$ characterising the Dirac spinor
of the nucleon in neutron matter. Further details see Fig.
\protect{\ref{fig:mskma}}}
\end{figure}

Significant differences between these two methods, however, can be observed in
the case of isospin asymmetric matter, like e.g.~pure neutron matter. The
effective masses obtained from the single-particle spectrum are significantly
larger than the range of $M^*(k)$, which are derived at the same density from
the projection method (see Fig.~\ref{fig:msnma}). This confirms the observation
of Ulrych et al.~\cite{ulr1} who studied a parametrization of $G$ published by
Boersma and Malfliet\cite{boer1}. This demonstrates that the analysis of the the
single-particle energy is not a reliable tool to determine the Dirac structure
of the self-energy: accidentally it works well for symmetric nuclear matter
for the NN interactions considered, it fails, however, for asymmetric nuclear
matter.

\begin{figure}[tb]
\begin{center}
\epsfig{file=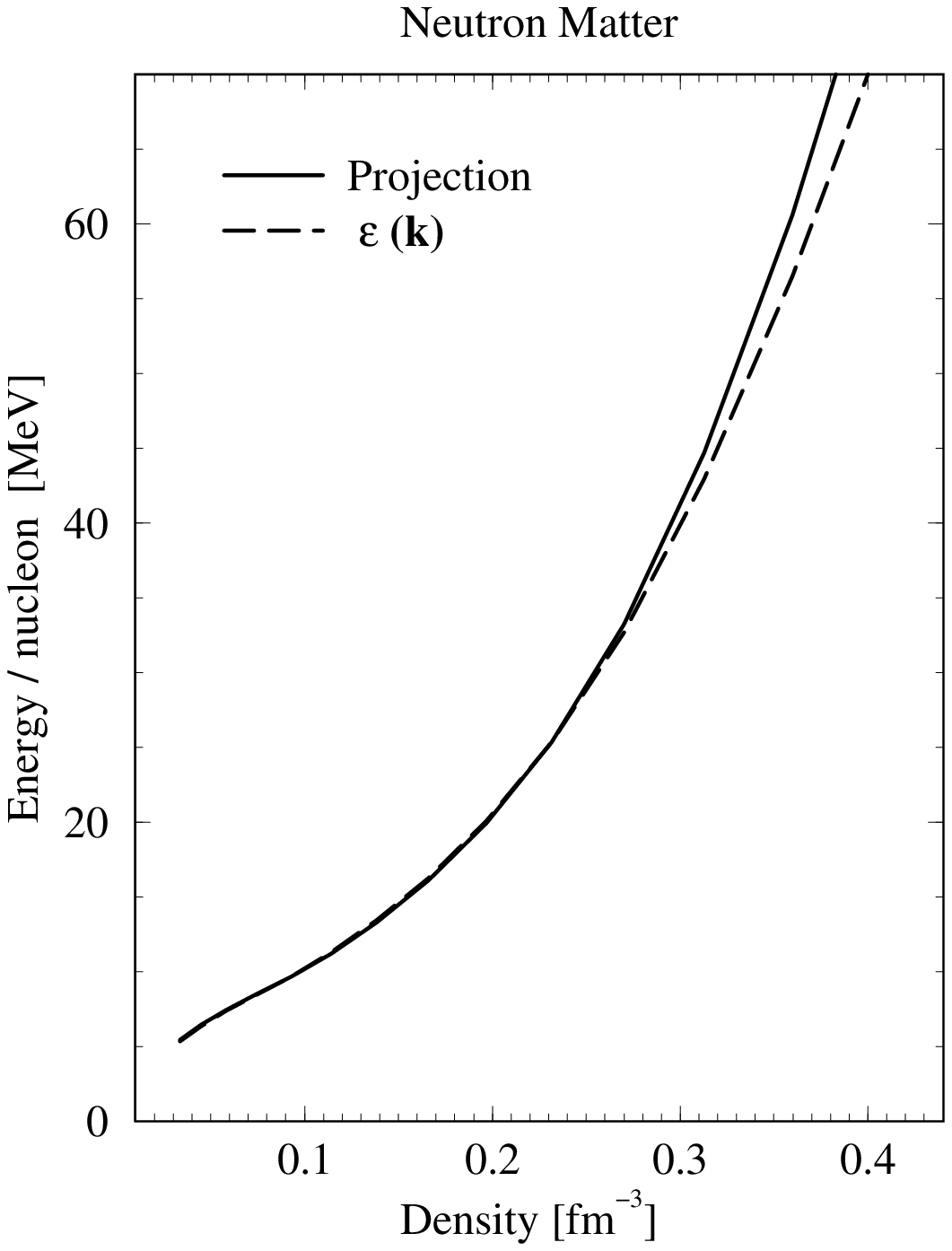,width=6cm}
\end{center}
\caption{\label{fig:ebnm} Calculated binding energy of neutron matter as a
function of the density using the Bonn A potential. The results derived from the
projection method are compared to those derived from the analysis of $\epsilon
(k)$.}
\end{figure}

How does this affect the calculated binding energies. As an example results for
the binding energy of pure neutron matter calculated for the Bonn A potential
are
presented in Fig.~\ref{fig:ebnm}. The two methods yield almost identical results
at low densities. At densities around 0.4 fm$^{-3}$, which corresponds 2.5 times
the saturation density of nuclear matter, the evaluation based on the simple
analysis of $\epsilon(k)$ underestimates the energy by 10 percent. 

\begin{figure}[tb]
\begin{center}
\epsfig{file=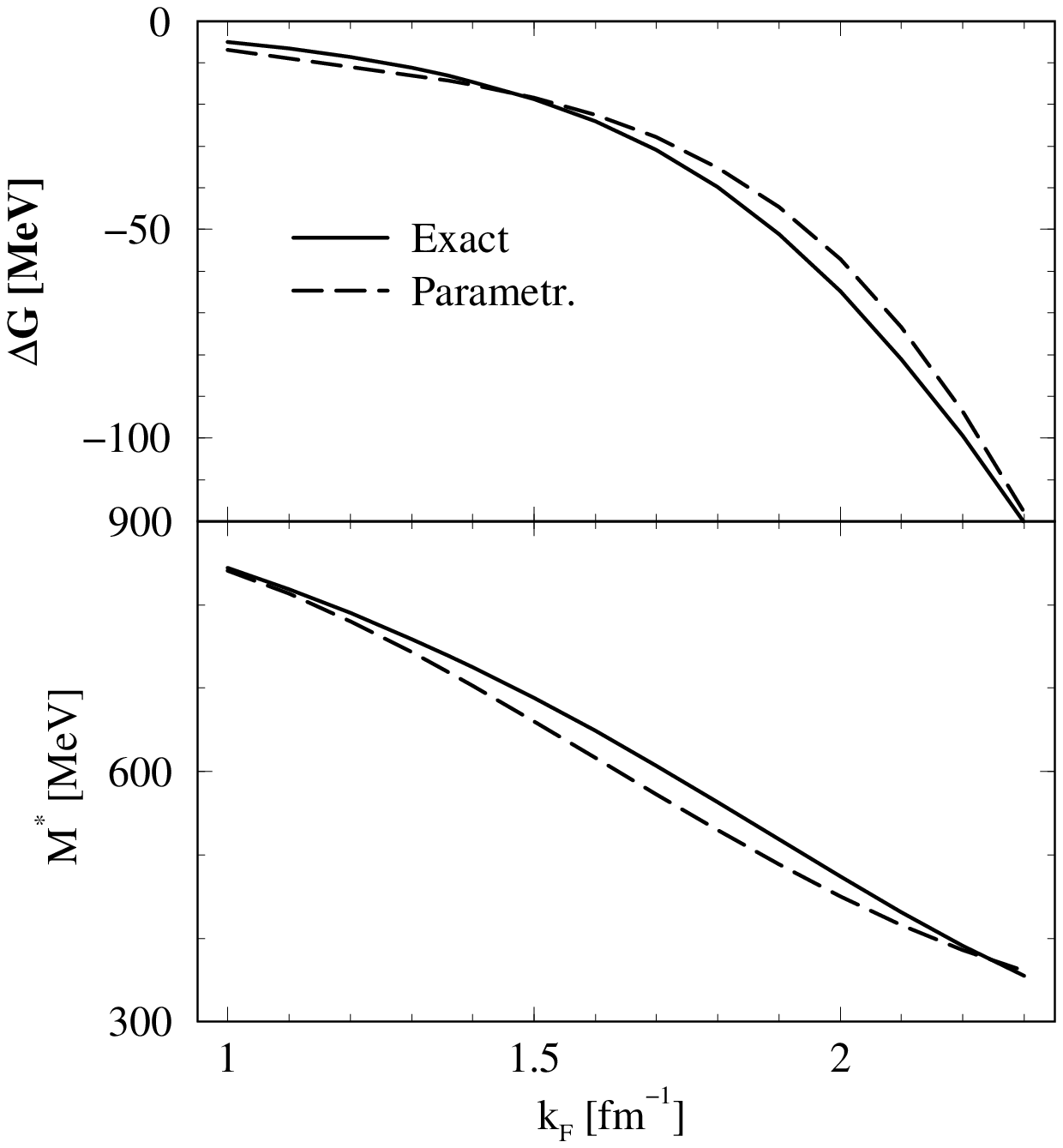,width=10cm}
\end{center}
\caption{\label{fig:para_comp} Contribution of $\Delta G$ to the energy of
neutron matter and the effective mass $M^*$, calculated at $k=k_F$, as a
function of $k_F$. The results of the direct calculation (solid lines) are
compared to the prediction of the parametrization of $\Delta G$ according 
to eq.(\protect\ref{eq:pardelg}) assuming that this parametrization is
independent of the isospin asymmetry.}
\end{figure}

Finally, we add a remark on the asymmetry dependence of the parametrization of
$\Delta G$ according to (\ref{eq:pardelg}). If one employs the density dependent
parameters for $\Delta G$ derived in symmetric nuclear as displayed in
Fig.~\ref{fig:coudG}, assuming that the parametrization of $\Delta G$ is
independent of the isospin asymmetry, also for pure neutron matter one obtains
effective masses ($M^*$ at $k=k_F$) and contributions of $\Delta G$ to the
energy per nucleon in neutron matter as presented by the dashed lines in
Fig.~\ref{fig:para_comp}. These results are very close to the corresponding
values derived from a direct determination of $\Delta G$ in neutron matter.
This indicates that the underlying assumption, the parametrization of $\Delta
G$ is independent of the isospin asymmetry is quite reasonable. Therefore one
may use the parametrization of $\Delta G$ as displayed in Fig.~\ref{fig:coudG}
for Dirac-Brueckner-Hartree-Fock studies of finite nuclei with $N=Z$ as well as
$N\not= Z$.    

\section{Conclusions}
A method is presented which determines the Dirac structure of the Brueckner 
$G$-matrix 
from its matrix elements between positive energy spinors only. The usual
projection method is model dependent as it depends on the choice of the
covariant operators. This model dependence is demonstrated for various
meson-exchange terms. In order
to minimise the model-dependence of the projection method $G$ is split into
the bare interaction $V$ and the correction $\Delta G$ reflecting the effects of
NN correlations. Since the Dirac structure of $V$ is known, the projection
methods is applied to the correction $\Delta G$ only. This analysis allows an
explicit study of the correlation effects on the Dirac structure of the nucleon
self-energy. A simple parametrization of $\Delta G$ and its density dependence
is presented in terms of the exchange of pseudo mesons with infinite mass. This
parametrization could be useful for the study of finite nuclei. 
It is shown that the simple method, which determines the Dirac
structure of the self-energy from the momentum dependence of the single-particle
energy yields fairly good results for symmetric nuclear matter but fails for
asymmetric matter.

This investigation has been supported by the SFB 382 of the ``Deutsche
Forschungsgemeinschaft''.

\end{document}